\documentclass[11pt]{article}
\usepackage{moriond,epsfig}
\usepackage{bigcenter}

\def\Journal#1#2#3#4{{#1} {\bf #2}, #3 (#4)}

\def\NPB{{\em Nucl. Phys.} B}

\def\PRL{\em Phys. Rev. Lett.}
\def\PRD{{\em Phys. Rev.} D}

\def\PL{{\em Phys. Lett.}}
\def\PR{{\em Phys. Rept.}}
\def\EPJC{{\em Eur. Phys. J.} C}
\def\JHEP{{\em JHEP}}

\def\be{\begin{equation}}
\def\ee{\end{equation}}
\def\bea{\begin{eqnarray}}
\def\eea{\end{eqnarray}}

\newcommand{\gsim}{\raisebox{-0.13cm}{~\shortstack{$>$ \\[-0.07cm] $\sim$}}~}
\begin{document}
\vspace*{4cm}
\title{THE SUPERSYMMETRIC HIGGS BOUNDS\\
 AT THE TEVATRON AND THE LHC}

\author{Julien Baglio}

\address{Laboratoire de Physique Th\'eorique\\
Universit\'e Paris-Sud 11 and CNRS,\\
Bat. 210, 91405 Orsay Cedex, France}

\maketitle\abstracts{
MSSM Higgs bosons are the most promising way to discover Higgs physics
at hadronic colliders since their cross section is enhanced compared
to that of the Standard Model. We will present theoretical predictions for their
production and decay in the Higgs$\to \tau \tau$ channel at the
Tevatron and the LHC, focusing on the theoretical uncertainties that
affect them. The inferred SUSY Higgs bounds on
the $[\tan \beta ; M_A ]$ plane and the impact of these uncertainties
will also be discussed.}

\section{Introduction}

\indent
The search for the Higgs bosons which are a trace of the electroweak
symmetry breaking\cite{Higgs:1964ia,Englert:1964et} is the main goal
for current high--energy
colliders. In the minimal supersymmetric extension of the Standard
Model (SM), one of the most attractive solutions of the hierarchy
problem in the SM\cite{Djouadi:2005gj}, two Higgs doublets are
required to cancel anomalies, which then lead to five Higgs states:
the CP--even $h,H$, the CP--odd $A$ and the two charged Higgs bosons $H^{\pm}$.

At tree--level two parameters in the Minimal Supersymmetric Standard
Model (MSSM) describe the Higgs sector: the vacuum
expectation values (vev) ratio between the two Higgs doublets
$\displaystyle \tan \beta = \frac{v_1}{v_2}$ and the CP--odd Higgs
mass $M_A$.

At high $\tan \beta$ values, $\tan \beta \gsim 10$, either $h$ or $H$
is SM--like and its couplings to other particles are the same as those
of the SM Higgs boson, while the other CP--even state behaves as
the CP--odd $A$: same couplings and almost same mass. We will
denote these two states as $\Phi$ in the next sections. This behaviour
occurs in current MSSM Higgs
benchmarks scenarios\cite{Carena:2002qg} which are considered at the
Fermilab Tevatron\cite{Benjamin:2010xb} and the CERN
LHC\cite{ATLAS-CONF-2011-024,Chatrchyan:2011nx} colliders.

$b$--processes are dominant as they are proportionnal to $\tan \beta$
contrary to that of the top--loop. We will thus consider the
gluon--gluon fusion Higgs production through bottom quark
loop\cite{Georgi:1977gs,Spira:1995rr} and the $b\bar b$ fusion
channel\cite{Dicus:1988cx,Campbell:2002zm,Maltoni:2003pn,Harlander:2003ai}, followed
by the Higgs$\to \tau^+ \tau^{-}$
desintegration. Squark loops can be safely neglected while SUSY
$\Delta_b$ corrections to the $\Phi b \bar
b$ coupling nearly cancel out in the production cross section times
branching ratio calculation\cite{Baglio:2011xz}.

We will present numerical results at the Tevatron and the lHC (LHC at
7 TeV) for $\tan \beta = 1$, which means that we have to multiply by
$2 \tan^2 \beta$ for actual values. Theoretical uncertainties will also
be presented and their implications on the MSSM parameter space
limits will be discussed. A more detailed discussion can be found in
Refs.\cite{Baglio:2011xz,Baglio:2010bn,Baglio:2010ae}

\section{SUSY Neutral Higgs production at the Tevatron and the lHC}

\subsection{$gg\to \Phi$ channel}

The Higgs bosons in the gluon--gluon fusion channel is produced through top and bottom
quarks loops. At $\tan \beta \gsim 10$ values the
top loop is strongly suppressed because $\Phi t \bar t$ is inversely proportionnal
to $\tan \beta$ contrary to the $\Phi b \bar b$ coupling. Although the
top loop is known up to next--to-next--to--leading order (NNLO) in
QCD, the $b$--loop is known up to next--to--leading order (NLO)
only\cite{Spira:1995rr}. We will use NLO MSTW 2008 parton distribution
functions (PDF) set\cite{Martin:2009iq}. We consider the standard QCD theoretical
uncertainties that have been
discussed in Refs.\cite{Baglio:2011xz,Baglio:2010bn,Baglio:2010ae}.

It is customary to estimate the uncertainty due to the missing higher
order terms in a perturbative calculation by varying the
renormalisation scale $\mu_R$ and the factorisation scale $\mu_F$
around a central scale $\mu_0$: $\frac{\mu_0}{\kappa}\leq
\mu_R,\mu_F\leq \kappa \mu_0$. We take $\mu_0=\frac{1}{2} M_\Phi$ as
the central scale in order to be consistent with the SM
calculation\cite{Baglio:2010um}, as one of the
CP--even Higgs is SM--like. We use $\kappa=2$ in the gluon--gluon
fusion channel and obtain $\Delta\sigma/\sigma \simeq \pm 20\%$ at the
Tevatron ($\pm 15\%$ at the lHC).

The next source of uncertainties is due to the combined uncertainty
from the PDF and $\alpha_S$ coupling. We use MSTW
collaboration scheme\cite{Martin:2009bu} and calculate the PDF+$\Delta^{\rm
  exp+th}\alpha_S$ 90\% CL uncertainty which is equivalent to the MSTW
PDF4LHC recommandation\cite{PDF4LHC} and we obtain $\Delta\sigma/\sigma
\simeq \pm 10\%$ both at the Tevatron and the lHC.

The last important uncertainty is specific to the MSSM case, and deals
with the $m_b$ mass. There are two types of uncertainties:
 the experimental errors on the $\overline{\rm MS}$ $\bar m_b (\bar m_b)$
value and the uncertainty due to the scheme choice for the
renormalisation of the $b$--mass. The first uncertainty will cancel
out in the production cross section times branching ratio (see below)
but not the other one. We obtain $\Delta \sigma/\sigma \simeq \pm
15\%$ at both colliders due to these $b$--quark issues.

\begin{figure}
\begin{bigcenter}

\epsfig{figure=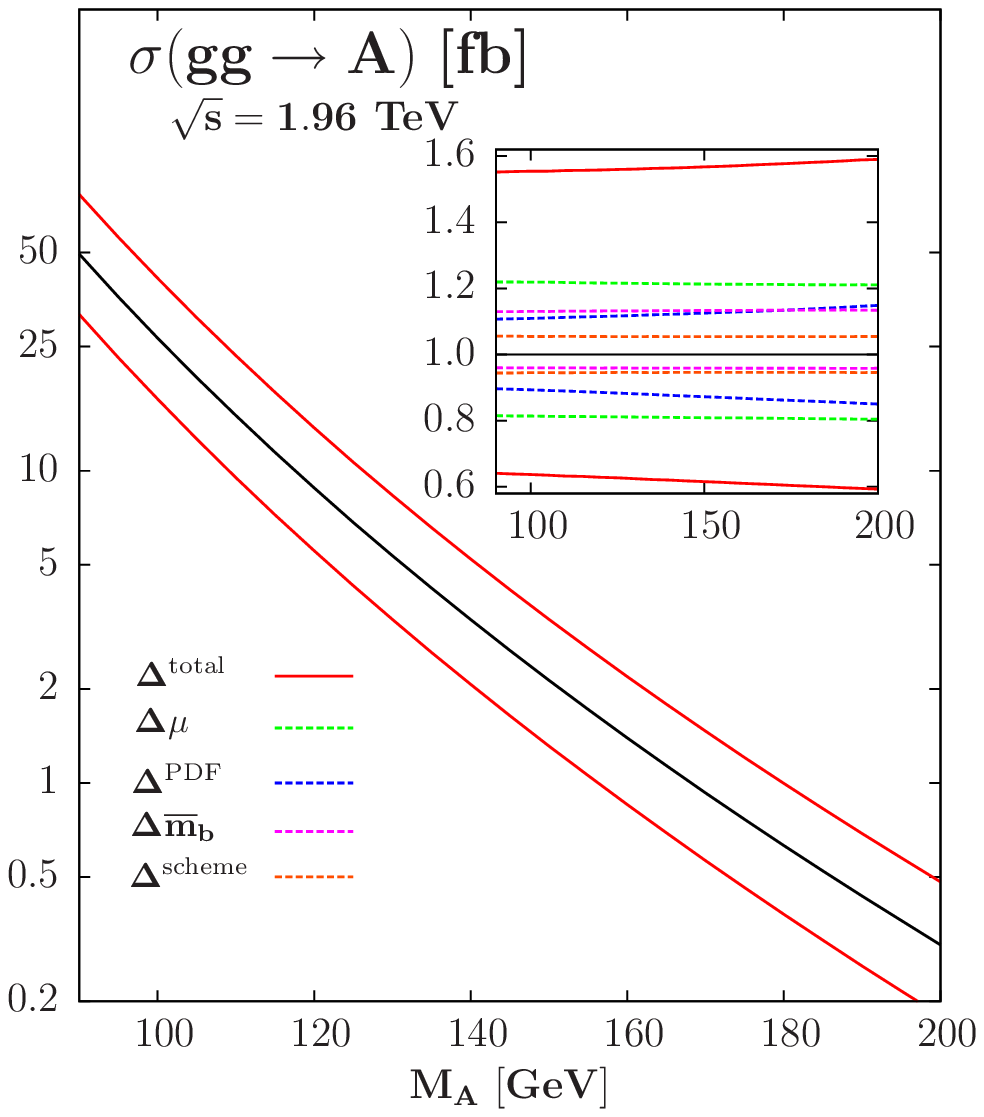,scale=0.45}
\epsfig{figure=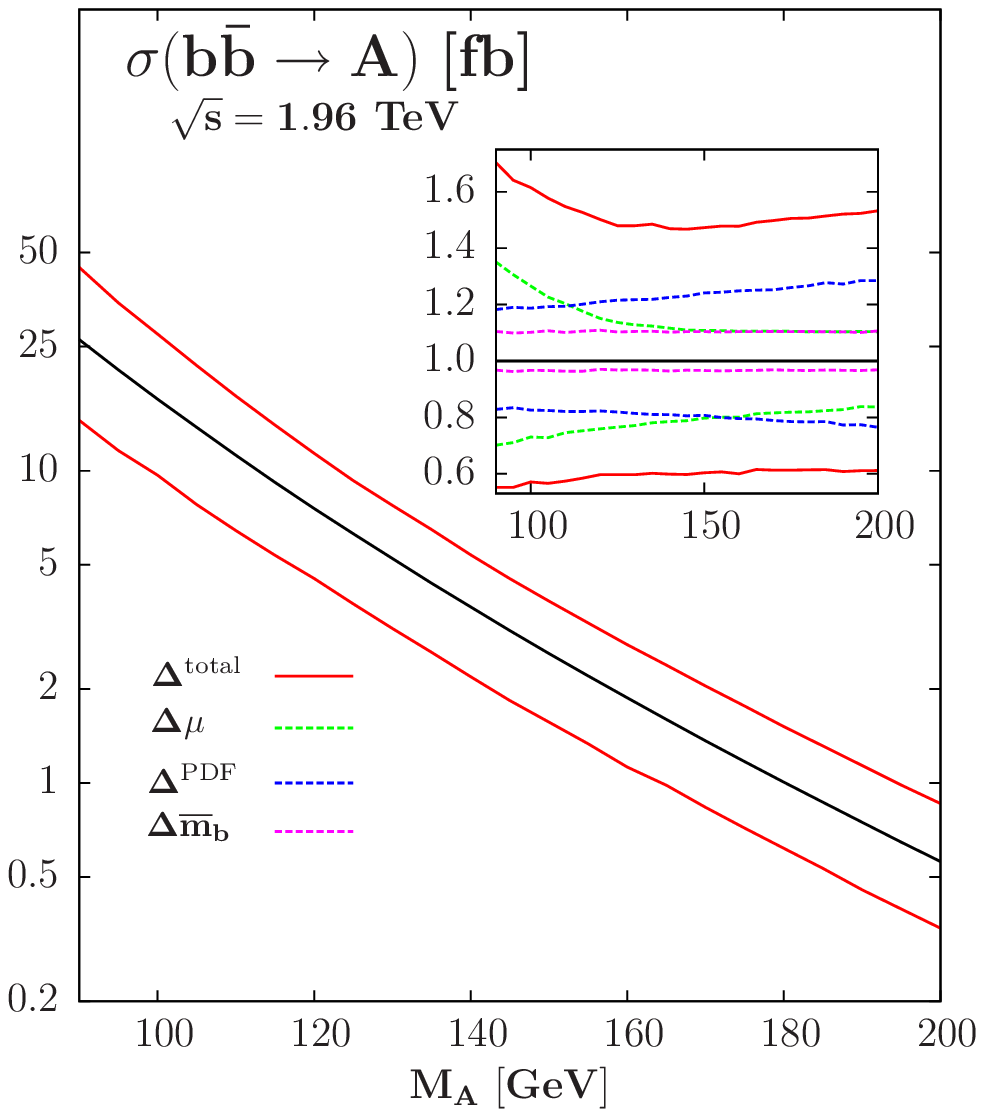,scale=0.45}
\epsfig{figure=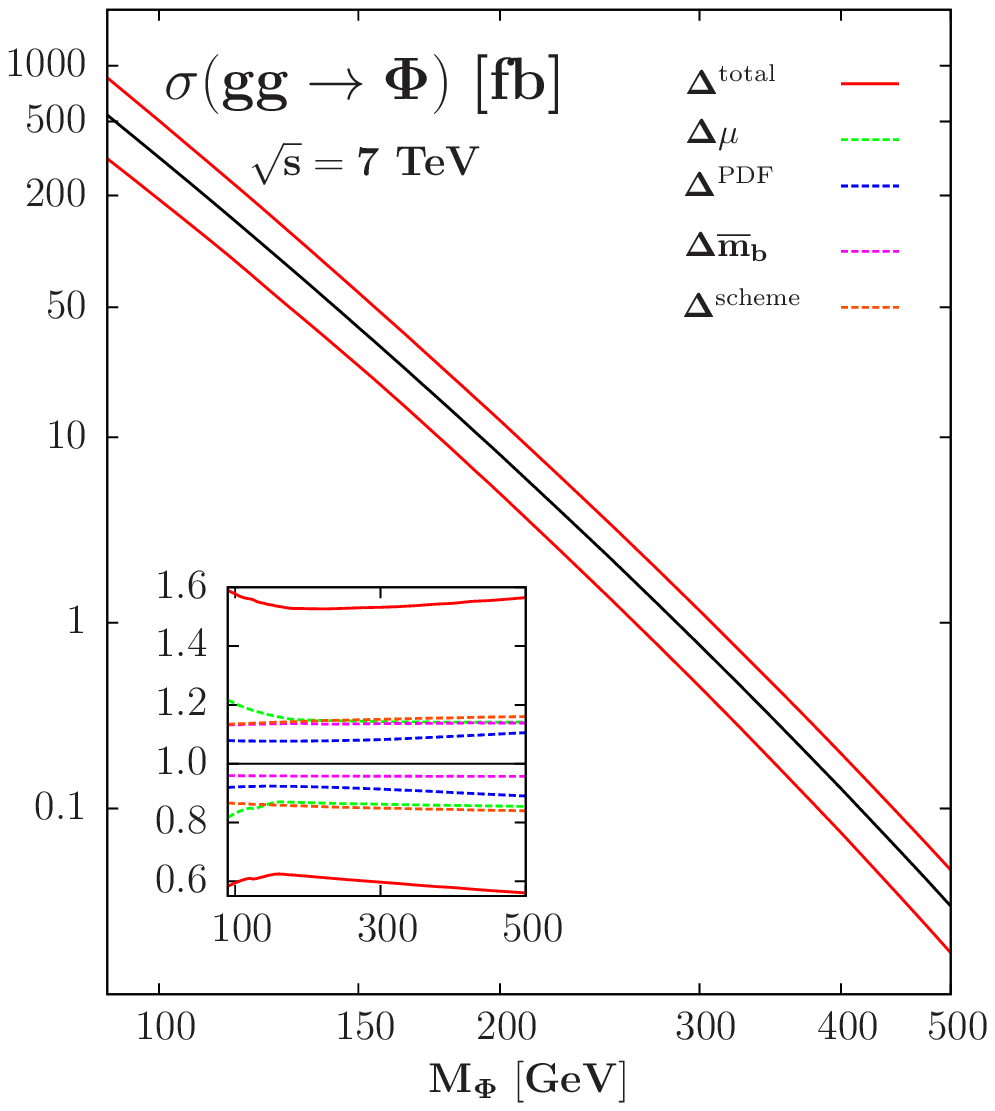,scale=0.45}
\epsfig{figure=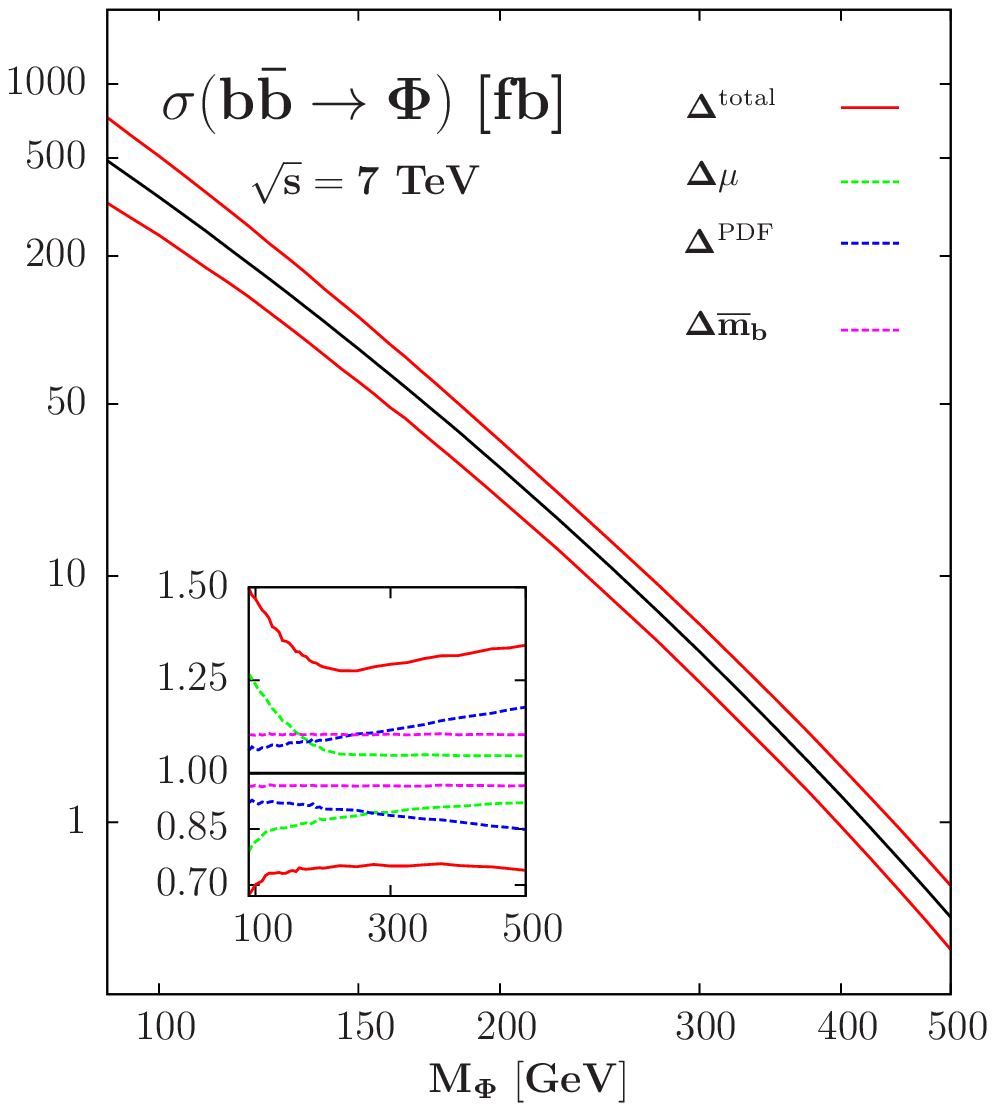,scale=0.45}

\caption{$\sigma^{\rm NLO}_{gg\to \Phi}$ and $\sigma^{\rm NNLO}_{b
    \bar b\to \Phi}$ central cross sections using
  MSTW 2008 PDFs and $\Phi b \bar b$ unit couplings together with total
  uncertainties at the Tevatron (left) and at the lHC
  (right). In the insert are shown the individual sources of
  uncertainties normalised to the central cross section.
\label{fig:ggPhi}}
\end{bigcenter}
\end{figure}

All these individual sources of uncertainties are shown in
Fig.\ref{fig:ggPhi} in the insert. We also display the total
uncertainty on the cross section when combining the uncertainties
according to the procedure developed in Ref.\cite{Baglio:2010bn}. We
obtain $\Delta \sigma/\sigma \simeq +58\%,-40\%$ at the Tevatron and
$\Delta \sigma/\sigma \simeq +53\%,-38\%$ at the lHC.

\subsection{$b\bar b \to \Phi$ channel}

The bottom quark fusion channel is strongly enhanced because of the
$\tan \beta$ effect in $b$--quark processes. This channel is known in the
SM up to NNLO in QCD\cite{Harlander:2003ai} and we rescale the
predictions with the MSSM $\Phi b \bar b$ coupling to obtain a NNLO MSSM
prediction. We use the same PDF set as for gluon--gluon fusion and
consider the same set of theoretical uncertainties.

For the scale uncertainties we consider here $\kappa=3$ instead of
$\kappa=2$ in $g g\to \Phi$. Indeed this is well known that either
four or five active flavours schemes can be used for the
calculation. The two predictions differ
significantly\cite{Assamagan:2004mu} and one way to reconcile them is
to allow such a scale interval. Furthermore this also allows the
inclusion of the $b$--mass scheme uncertainty that was obtained separately in the
gluon--gluon fusion calculation. We obtain in the end $\Delta \sigma/\sigma \simeq 30\%$
for low masses at the Tevatron ($\pm 25\%$ at the lHC).

The combined PDF+$\alpha_S$ uncertainty is calculated exactly as in
the gluon--gluon fusion case. We obtain in the bottom quark fusion
$\Delta\sigma/\sigma \simeq \pm 20\%$ for low masses and $\simeq \pm
30\%$ for high masses at the Tevatron ($\simeq 10\%$ at low masses and
$\simeq \pm 20\%$ at high masses at the lHC).

The last uncertainty involves only the experimental $b$--mass error. We
obtain a $+10\%,-4\%$ uncertainty at the Tevatron (nearly the same at
the lHC), which as discussed in the next section will cancel out in the cross section
times branching ratio calculation.


All the uncertainties are displayed in Fig.\ref{fig:ggPhi}. The
overall total uncertainty is $\Delta \sigma/\sigma \simeq +50\%,-40\%$
at the Tevatron ($+40\%, -30\%$ at the lHC).

\subsection{Combinaison with the $\Phi\to \tau \tau$ branching ratio}

We finally evaluate the combinaison of the two production channels
together with the branching ratio $\Phi\to \tau^+ \tau^-$. The issue
is how to combine the uncertainties and we proceed as stated in
Refs.\cite{Baglio:2010ae,Baglio:2010bn}: the cross section uncertainties are
weighted according to their importance and we add linearly the decay
branching ratio uncertainty which is $\simeq +4\%,-9\%$ on BR$(\Phi
\to \tau^+ \tau^- \simeq 10\%)$\cite{Baglio:2010ae}. In this
procedure, as the uncertainties due to the experimental errors on
$b$--mass are anti-correlated in the production and decay, they
cancel out.

\begin{figure}
\begin{center}
\epsfig{figure=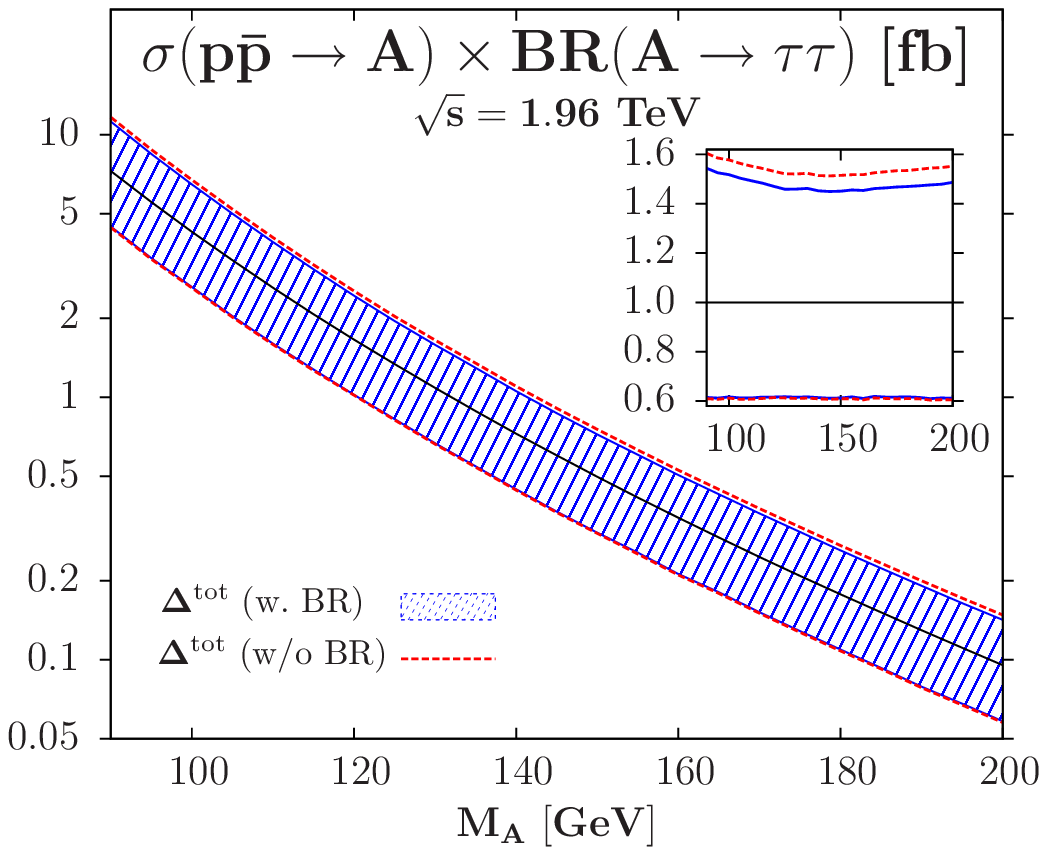,scale=0.48}\hspace{2mm}
\epsfig{figure=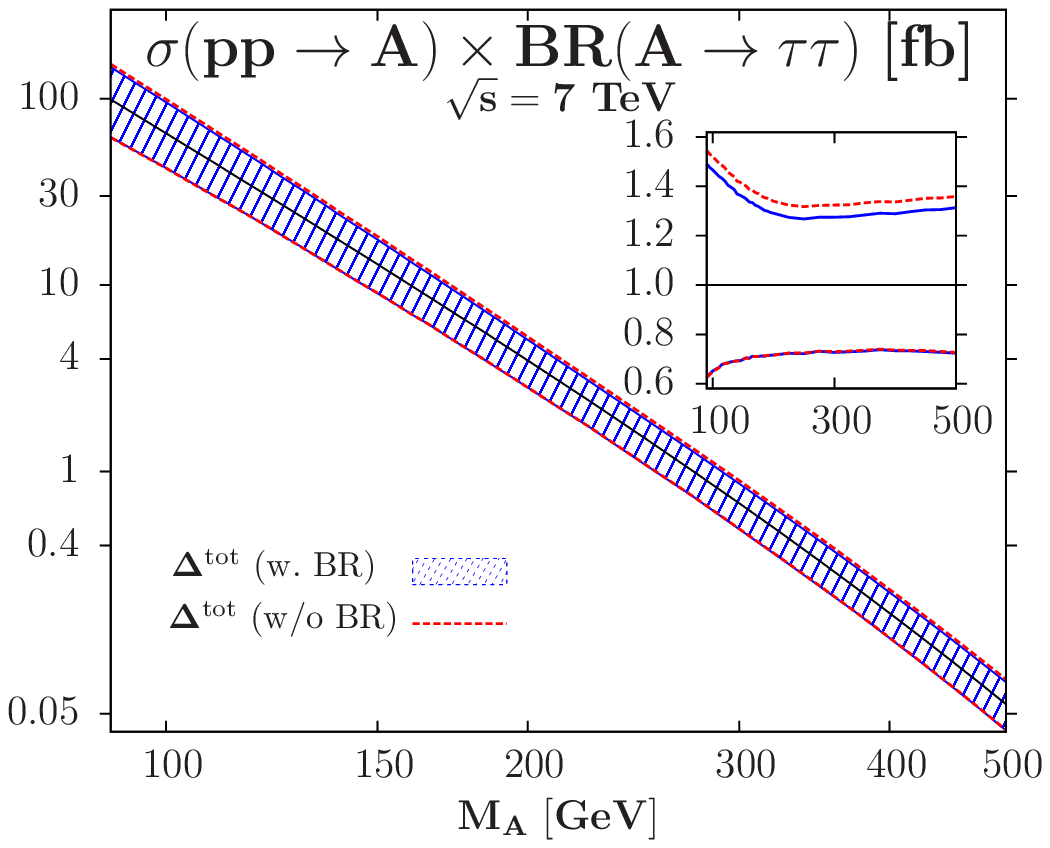,scale=0.48}
\caption{$\sigma(p\bar p\to A)\times {\rm BR}(A\to \tau^+ \tau^-)$
  as a function of $M_A$ at the Tevatron (left) and at the lHC
  (right), together with the associated overall theoretical
  uncertainty; the uncertainty when excluding that on the branching
  ratio is also displayed. In the inserts, shown are the relative
  deviations from the central values.
\label{fig:phitautau}}
\end{center}
\end{figure}

We then obtain $\Delta (\sigma \times {\rm BR}) / (\sigma \times {\rm
  BR}) \simeq +50\%,-39\%$ at the Tevatron and $\simeq +35\%,-30\%$ at
the lHC, as shown in Fig.\ref{fig:phitautau}.

\section{Higgs bounds on the MSSM parameter space}

We are left to evaluate the impact of the theoretical uncertainties
calculated above on the 95\% CL limits in the $[\tan \beta ; M_A]$ plane using
the experimental results at the Tevatron and the lHC. The results
presented above are quite model independant as they do not depend on
the details of the MSSM model as long as we have a degeneracy in the
$h/H,A$ spectrum. We apply the limit on the minimal cross section
times branching ratio instead of the central prediction in order to
take into account the theoretical uncertainties.

\begin{figure}
\begin{center}
\epsfig{figure=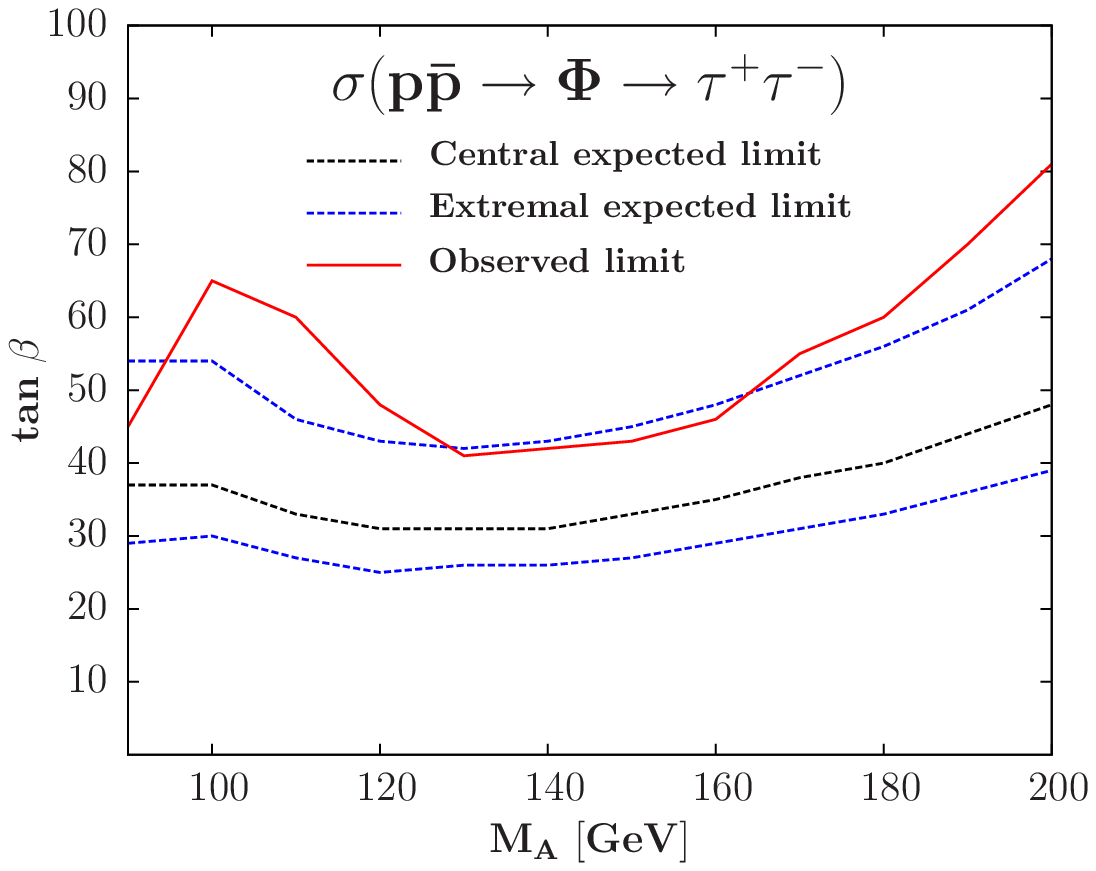,scale=0.48}\hspace{2mm}
\epsfig{figure=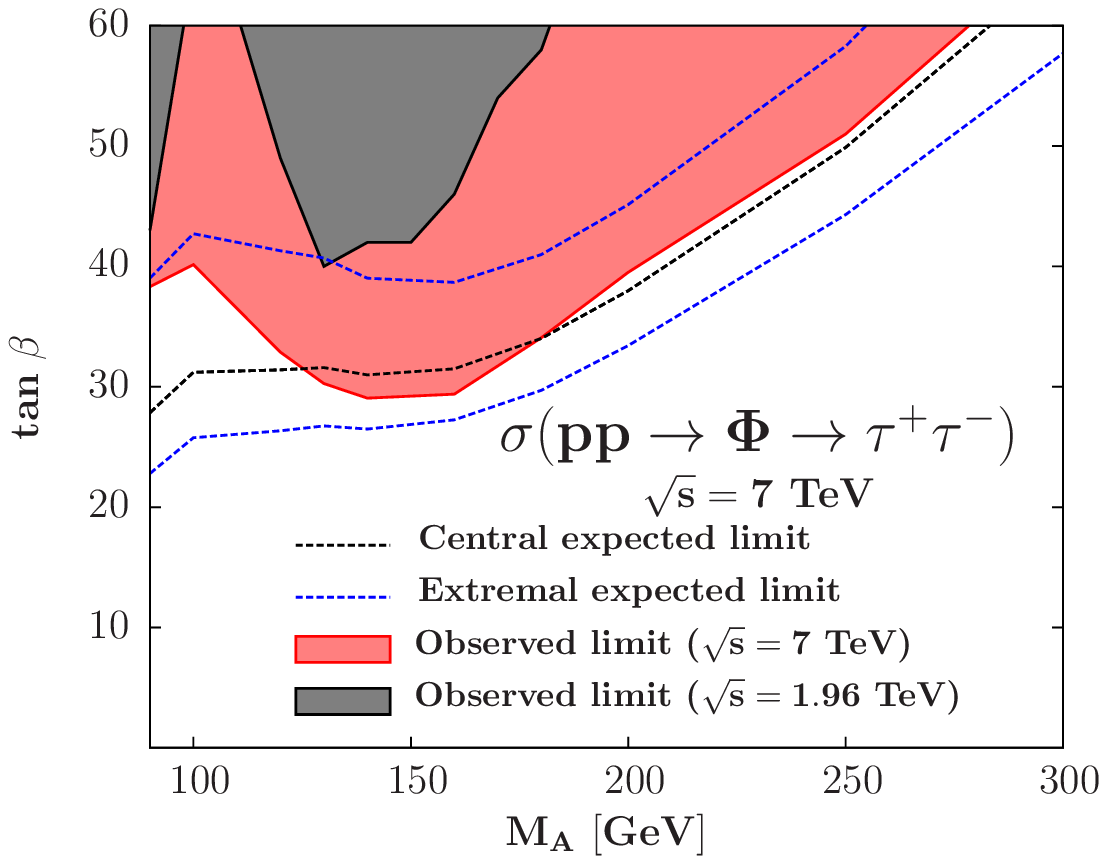,scale=0.48}
\caption{Contours for the expected $\sigma(p\bar p\to\Phi\to \tau^+
  \tau^-$ rate at the Tevatron (left) and at the lHC (right)
  in the $[M_A;\tan \beta]$ plane with the associated theory
  uncertainties, confronted to the 95\% CL exclusion limit obtained by
  CDF/D0 and CMS.
\label{fig:MSSMspace}}
\end{center}
\end{figure}

The result is shown in Fig.\ref{fig:MSSMspace} and the theoretical
uncertainties are extremely important. We obtain $\tan \beta >
45$ at the Tevatron, which thus reopens a large part of the parameter
space excluded by CDF/D0\cite{Benjamin:2010xb}. The comparaison with
CMS results at the lHC
shows a slight reduction of the exclusion limit as we obtain $\tan \beta >
29$ to be compared with $\tan \beta > 23$. The result is comparable to
what can be obtained with the theory uncertainty quoted by
CMS\cite{Chatrchyan:2011nx}.

\vspace{-3mm}
\section*{Acknowledgments}
J.B. would like to thank the Moriond 2011 organisers for the very
fruitful atmosphere and the organisation of the conference.


\end{document}